\documentclass{mn2e}
\usepackage{graphicx}
\usepackage{amsmath}
\usepackage{multirow}
\usepackage{hyperref}

\title[\textit{Chandra}/HETG view of NGC\,1365]{The \textit{Chandra}/HETG view of NGC\,1365~in a 
Compton-thick state}
\author[E. Nardini et al.]
{E.~Nardini,$^1$\thanks{E-mail: e.nardini@keele.ac.uk} J.~Gofford,$^{1,2}$ J.~N.~Reeves,$^{1,2}$ 
V.~Braito,$^3$ G.~Risaliti,$^4$ M.~Costa$^1$ \\
$^1$Astrophysics Group, School of Physical and Geographical Sciences, 
Keele University, Keele, Staffordshire ST5 5BG, UK \\
$^2$Department of Physics, University of Maryland Baltimore County, 
1000 Hilltop Circle, Baltimore, MD 21250, USA \\
$^3$INAF -- Osservatorio Astronomico di Brera, via E. Bianchi 46, I-23807 Merate, Italy \\
$^4$INAF -- Osservatorio Astrofisico di Arcetri, L.go E. Fermi 5, I-50125 Firenze, Italy}

\begin{document}


\pagerange{\pageref{firstpage}--\pageref{lastpage}} \pubyear{2015}

\maketitle

\label{firstpage}

\begin{abstract}
We present the analysis of a \textit{Chandra} High-Energy Transmission Grating 
(HETG) observation of the local Seyfert galaxy NGC\,1365. The source, well known 
for its dramatic X-ray spectral variability, was caught in a reflection-dominated, 
Compton-thick state. The high spatial resolution afforded by \textit{Chandra} 
allowed us to isolate the soft X-ray emission from the active nucleus, neglecting 
most of the contribution from the kpc-scale starburst ring. The HETG spectra thus 
revealed a wealth of He- and H-like lines from photoionized gas, whereas in larger 
aperture observations these are almost exclusively produced through collisional 
ionization in the circumnuclear environment. Once the residual thermal component 
is accounted for, the emission-line properties of the photoionized region close 
to the hard X-ray continuum source indicate that NGC\,1365 has some similarities 
to the local population of obscured active galaxies. In spite of the limited 
overall data quality, several soft X-ray lines seem to have fairly broad profiles 
($\sim$800--1300 km s$^{-1}$ full-width at half maximum), and a range of outflow 
velocities (up to $\sim$1600 km s$^{-1}$, but possibly reaching a few thousands 
km s$^{-1}$) appears to be involved. At higher energies, the K$\alpha$ fluorescence 
line from neutral iron is resolved with $> 99$ per cent confidence, and its width 
of $\sim$3000 km s$^{-1}$ points to an origin from the same broad-line region 
clouds responsible for eclipsing the X-ray source and likely shielding the 
narrow-line region.

\end{abstract}


\section{Introduction}

Irrespective of their different optical properties and classification, the energy source 
of active galactic nuclei (AGN) is strongly suggested to be intrinsically the same (e.g. 
Antonucci \& Miller 1985). When broad lines are lacking, a direct view of central engine 
would be prevented by a dusty absorber, covering a substantial, yet not complete fraction 
of the solid angle, and thus commonly envisaged as a toroidal structure (Krolik \& Begelman 
1988). The orientation-based idea that underlies AGN unification models (Antonucci 1993, 
and references therein) is still broadly accepted. While it was argued early on that the 
torus might well consist of individual \textit{clouds}, the multiple pieces of observational 
evidence accumulated in the last decade imply that the circumnuclear environment in AGN is 
characterized by a large degree of clumpiness and inhomogeneity on various physical scales, 
from tens of pc down to a few hundreds of gravitational radii ($r_\rmn{g} = GM_\rmn{BH}/c^2$) 
from the supermassive black hole (see Bianchi, Maiolino \& Risaliti 2012; Netzer 2015 for 
recent overviews). 

The nearby Seyfert galaxy NGC\,1365 ($z \simeq 0.0055$) is an ideal object to probe the 
complexity of nuclear obscuration in AGN. Optically classified as a type 1.8 (Maiolino 
\& Rieke 1995), NGC\,1365 has been extensively studied in the X-rays for its exceptional 
variability, which makes it the most striking example of a changing-look AGN (Matt, 
Guainazzi \& Maiolino 2003). Apparently due to a favourable line of sight, the inner 
accretion disc/X-ray corona system is seen through an erratic absorbing medium, whose 
marked column density gradients are betrayed by the random alternation of Compton-thin 
($N_\rmn{H} < 10^{23}$ cm$^{-2}$) and Compton-thick ($N_\rmn{H} > 10^{24}$ cm$^{-2}$) 
states. The switch from transmission- to reflection-dominated spectra (and back) can 
take less than two days, involving the transit of a single cloud (Risaliti et al. 2007). 
This kind of eclipses not only sets a tight empirical limit to the size of the X-ray 
source (no larger than a few tens of $r_\rmn{g}$), but also provides an implicit 
measure of the distance and density of the intervening blobs (both indicative of the 
gas in the Broad Line Region; BLR), and reveals unique details about their shape (e.g. 
Maiolino et al. 2010). Besides this, in the Compton-thin regime NGC\,1365 regularly 
exhibits four strong absorption lines in the $\sim$6.7--8.3 keV band. Their spacing 
prompts an identification with the K$\alpha$ and K$\beta$ pairs from 
Fe\,\textsc{xxv}--\textsc{xxvi}, blueshifted by reason of an outflow velocity of 
$v_\rmn{out} \sim 1000$--5000 km s$^{-1}$, variable over timescales of weeks/months 
(Risaliti et al. 2005; Nardini et al., in preparation). These lines could just trace 
the high-ionization phase of a much more pervasive wind, possibly including all the 
manifest absorption components (Connolly, McHardy \& Dwelly 2014; Braito et al. 2014). 

Adding to the overall merit of NGC\,1365 is one of the largest infrared luminosities 
($L_\rmn{IR} \sim 10^{11}\,L_{\sun}$) among nearby, non-interacting galaxies. This 
mostly arises from the great number of star-forming clusters in the central $\sim$2--3 
kpc, disposed in an elongated ring around the nucleus (Alonso-Herrero et al. 2012, and 
references therein). The powerful starburst is also the main source of the observed soft 
X-ray emission ($E < 3$ keV), which is found to be constant in flux and shape. Indeed, 
the $\sim$500-ks high-resolution spectrum obtained by combining all the 2004 and 2007 
data from the \textit{XMM--Newton} Reflection Grating Spectrometer (RGS) is rife with 
collisionally ionized lines from the hot diffuse gas, at odds with a typical obscured 
Seyfert (Guainazzi et al. 2009). In this paper we report on two \textit{Chandra} 
observations of NGC\,1365 taken with the High-Energy Transmission Grating (HETG; 
Canizares et al. 2005) over a span of four days, during which the AGN remained in 
a heavily absorbed, Compton-thick state. 

\section{Observations and Data Reduction}

NGC\,1365 was observed twice by \textit{Chandra} in 2012, on April 9 and three days 
later (ObsIDs 13920/1), for a total exposure of $\sim$200 ks. After a series of six 
15-ks snapshots in 2006 (Risaliti et al. 2007) and a previous one of the same length 
in 2002, these were the first observations of NGC\,1365 at high spectral resolution 
with \textit{Chandra}. The HETG consists of two sets of gratings, optimized for medium 
(MEG; 0.5--7 keV) and high (HEG; 0.8--10 keV) energies, respectively, and was used in 
combination with the Advanced CCD Imaging Spectrometer (ACIS-S) array. The data were 
reprocessed with the \textsc{ciao} version 4.4 software package and the v4.4.9 
Calibration Database (CALDB). For each observation, the MEG and HEG spectra were 
extracted from first-order events in both ($\pm 1$) diffraction arms. Redistribution 
matrices were created with the \textsc{ciao} tool \texttt{mkgrmf}, while effective 
area files were obtained through the \texttt{fullgarf} script. The spectra and detector 
responses from the $\pm 1$ orders were combined with the appropriate weights. Since the 
background is negligible, no correction was performed, and no systematic uncertainty 
was introduced. 

\begin{figure}
\includegraphics[width=8.5cm]{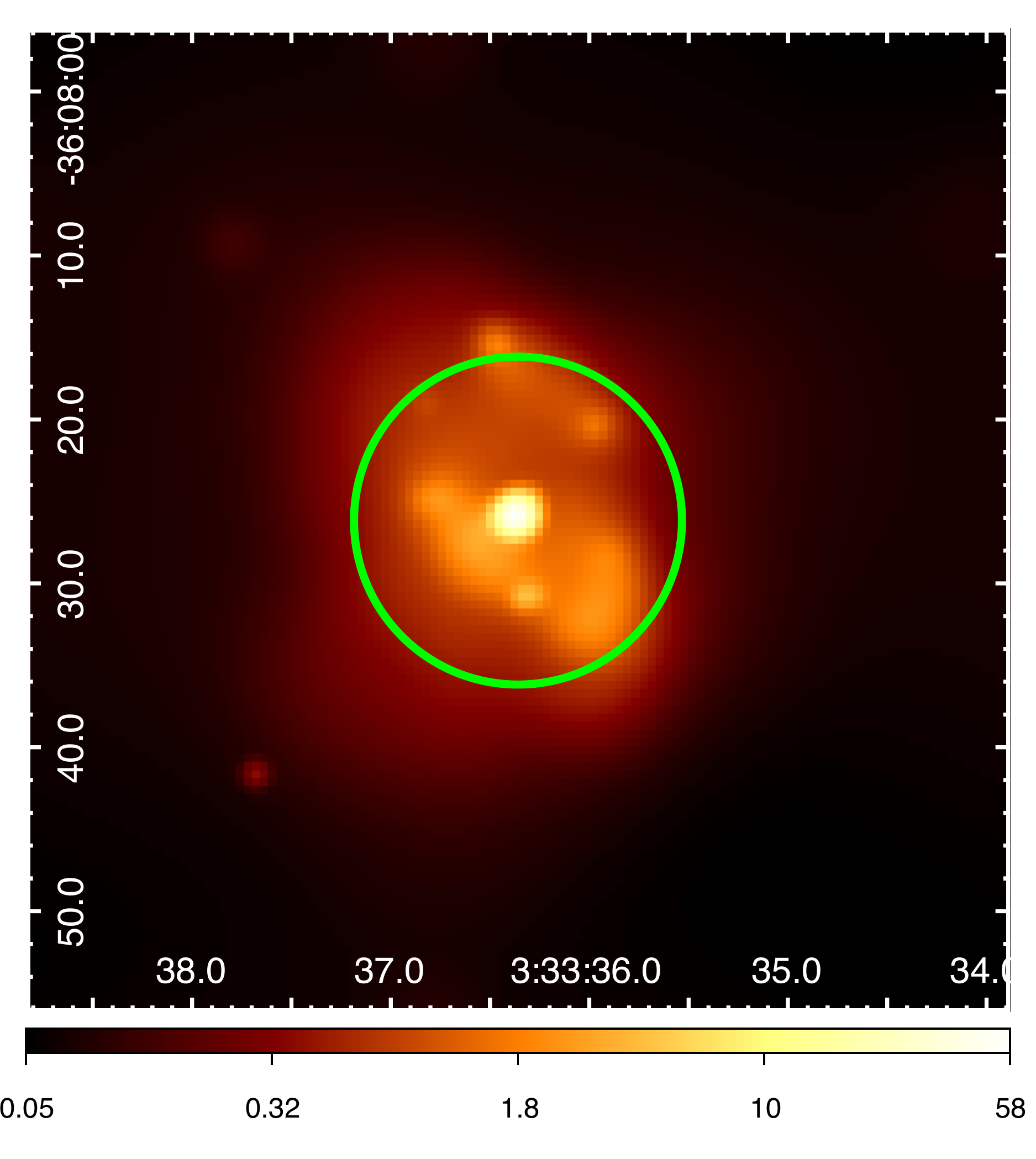}
\caption{X-ray image of the central $1\arcmin \times 1\arcmin$ region of NGC\,1365 
in the 0.3--3 keV energy range (from ObsIDs 13920/1), showing the extent of the diffuse 
starburst emission and the aperture adopted (green circle, 10$\arcsec$ radius). An 
adaptive smoothing has been applied following the same criteria of Wang et al. (2009).} 
\label{im}
\end{figure}

After verifying that the two observations are consistent with each other, in order 
to improve the data quality for the subsequent analysis we also merged the spectra 
(and auxiliary files) from ObsIDs 13920/1 into a single one. It is worth noting that 
we adopted a narrower extraction strip than the default one (20$\arcsec$ instead of 
35$\arcsec$): this allowed us to exclude part of the diffuse X-ray emission associated 
with the star-forming ring of $\sim$15$\arcsec$ diameter on kpc scale (1$\arcsec \sim 90$
 pc, Fig.~\ref{im}; see also Wang et al. 2009), and at the same time to extend the 
final HEG spectrum up to $\sim$8 keV. In all the spectral fits (carried out with 
\textsc{xspec} v12.8, with solar abundances from Wilms, Allen \& McCray 2000) we 
considered the 0.5--5 keV (MEG) and 1.5--8 keV (HEG) energy ranges, over which a 
total of $\sim$1580 and 1050 counts were collected, respectively. We initially binned 
the data to 2048 channels, corresponding to a $\Delta\lambda$ of 20 (MEG) and 
10 m\AA~(HEG), that is roughly the full width at half-maximum (FWHM) resolution. 
However, due to the limited number of counts, we further imposed a minimum of four 
counts per energy bin, and made use of the $C$-statistic (Cash 1979). Unless otherwise 
stated, the lines' energies are given in the rest frame of NGC\,1365, and all the 
uncertainties correspond to the 90\% confidence level for the single parameter of 
interest ($\Delta C = 2.71$). For simplicity, throughout this work we assumed a 
standard concordance cosmology with $H_0=70$ km s$^{-1}$ Mpc$^{-1}$, $\Omega_m=0.27$, 
and $\Omega_\Lambda=0.73$, although the luminosity distance entailed for NGC\,1365 
($\sim$21 Mpc) slightly differs from the one based on Cepheid variables of 
$D = 18.6 \pm 1.9$ Mpc (Madore et al. 1998).

\begin{figure}
\includegraphics[width=8.5cm]{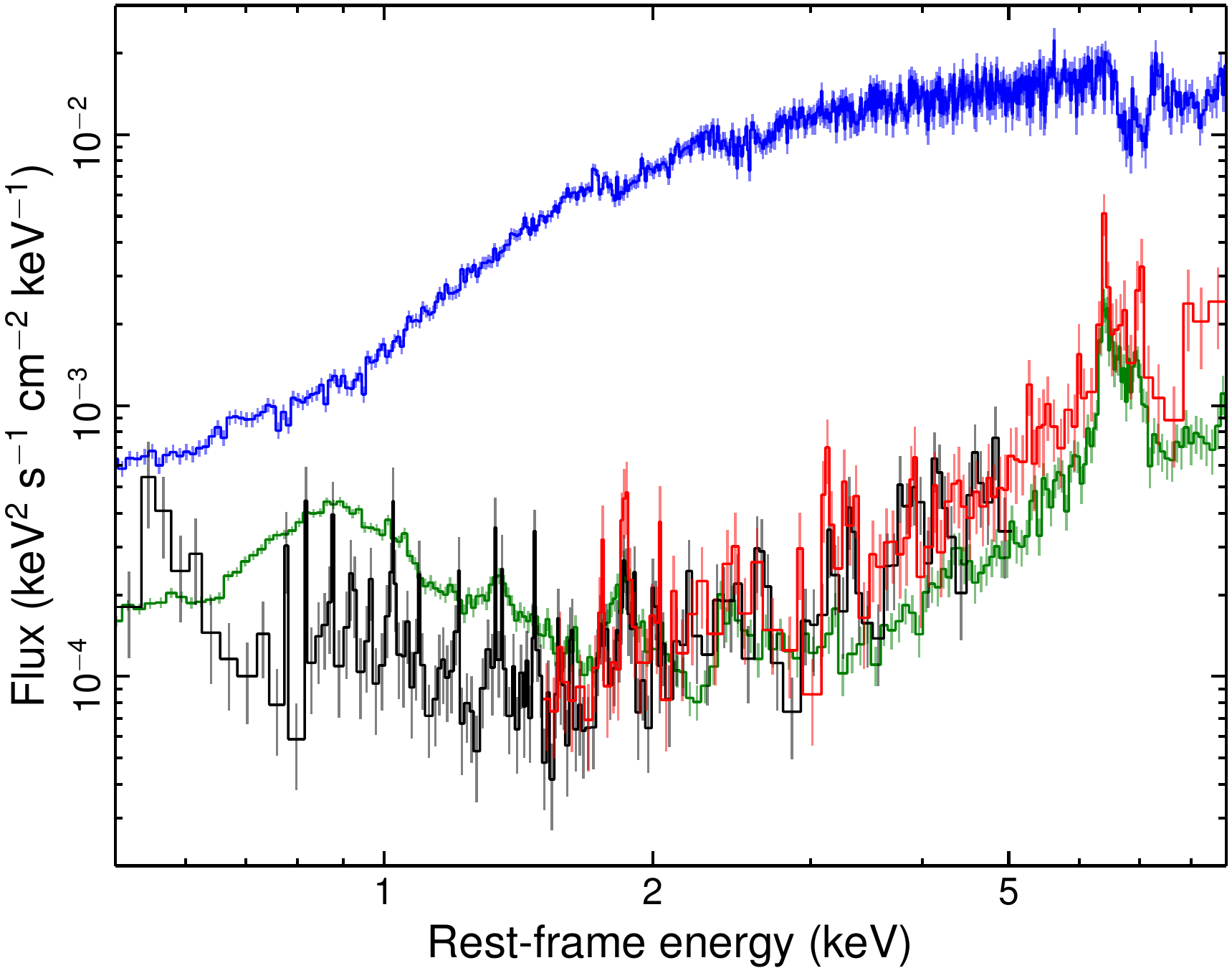}
\caption{\textit{Chandra} HETG spectra of NGC\,1365, plotted in black (MEG) 
and red (HEG), respectively. The historical extremes of transmission- and 
reflection-dominated states are also shown for comparison. The source has 
the typical Compton-thick spectral shape at 2--10 keV, and is $\sim$15 times 
dimmer than during the final part of the January 2013 \textit{XMM--Newton} 
observation (blue), when the neutral column density dropped to $N_\rmn{H} 
\sim 10^{22}$ cm$^{-2}$. Indeed, the flux lies above the minimum reached in 
the 2007 \textit{XMM--Newton} monitoring (green) just by a factor of $\sim$1.8. 
Note, however, that the narrower extraction region adopted for \textit{Chandra} 
allowed us to disentangle the nuclear soft X-ray component from most of the 
circumnuclear starburst emission. (All the data were rebinned for display 
purposes).} 
\label{xs}
\end{figure}

\section{Spectral Analysis}

The HETG spectra of NGC\,1365 are shown in Fig.~\ref{xs}, compared to the brightest flux 
state ever reached by the source during an \textit{XMM--Newton} observation performed only 
nine months later in January 2013, when the column density towards the AGN was just about 
$N_\rmn{H} \sim 10^{22}$ cm$^{-2}$ (Braito et al. 2014; Rivers et al. 2015). \textit{Chandra} 
clearly met a Compton-thick state, exhibiting a faint and very hard ($\Gamma = -1.5$) 
continuum above 3 keV, which closely resembles the lowest reflection-dominated spectrum 
of the first \textit{XMM--Newton} orbit in the 2007 five-day monitoring (Fig.~\ref{xs}; 
Risaliti et al. 2009a). In the following analysis, we first focus on the overall shape 
of the continuum, and then take into account the rich emission-line spectrum in the soft 
X-rays. 

\subsection{Baseline continuum}

As a starting point we adopted a model that simply consists of three different AGN-related 
components: the primary power-law continuum, attenuated by a foreground screen of cold gas 
at rest in the source frame ($z = 0.005457$); reflection off the irradiated matter, described 
as a \texttt{pexrav} template (Magdziarz \& Zdziarski 1995) with solar abundances, inclination 
of 45$\degr$, and variable strength; and a soft X-ray power law, which is required by the modest 
spectral rise below $\sim$1.5 keV and is intrepreted as the faint, scattered AGN continuum. 
The latter component can be fitted in several alternative ways, for instance as a thermal 
bremsstrahlung of $kT \sim 0.8$ keV, possibly associated with starburst-driven shocks. However, 
due to its negligible impact on the final results and to the small ($\sim$1--2 per cent) 
scattering efficiency implied, we retained for simplicity the link of this soft continuum 
with the AGN, thus assuming a common photon index for the various forms of reprocessed AGN 
emission (i.e. transmitted, reflected and scattered). This model (hereafter tagged as model 
A for brevity) and all the subsequent ones also include the Galactic column density (fixed 
to $N_\rmn{H} = 1.34 \times 10^{20}$ cm$^{-2}$; Kalberla et al. 2005) and the conspicuous 
iron fluorescence line around 6.4 keV, whose properties remain consistent throughout the 
fitting steps. 

\begin{table*}
\caption{Best-fit continuum parameters for the different models (A--E) and components: 
primary (1), scattered (2) and reflected (3) AGN emission, hot (4) and warm (5) thermal 
emission. $N_\rmn{H}$: foreground column density. $\Gamma$: power-law photon index. $A$: 
power-law normalization at 1 keV. $R$: reflection strength. $kT$: plasma temperature. 
$Z$: elemental abundance. EM: emission measure. (f) and (t) denote fixed and tied 
parameters. Each model also includes Galactic absorption and iron K$\alpha$ emission.}
\label{t1}
\begin{tabular}{l@{\hspace{30pt}}c@{\hspace{20pt}}c@{\hspace{20pt}}c@{\hspace{20pt}}c@{\hspace{20pt}}c}
\hline
Component & \multirow{2}{*}{Model A} & \multirow{2}{*}{Model B} & \multirow{2}{*}{Model C} & 
\multirow{2}{*}{Model D} & \multirow{2}{*}{Model E}\\[-0.5ex]
~~~~~~~~~~Parameter & & & & & \\
\hline
(1)~~\texttt{powerlaw}$_1$ & & & & & \\[0.5ex]
~~~~~~~~~~$N_\rmn{H}$ (10$^{22}$\,cm$^{-2}$) & 200(f) & 200(f) & 200(f) & 200(f) & 140$^{+19}_{-12}$\\[0.5ex]
~~~~~~~~~~$\Gamma$ & 2.0(f) & 2.0(f) & 2.0(f) & 2.0(f) & 2.10$^{+0.05}_{-0.06}$ \\[0.5ex]
~~~~~~~~~~$A$ (10$^{-4}$\,cm$^{-2}$\,s$^{-1}$\,keV$^{-1}$) & 77$^{+22}_{-20}$ & 60$^{+21}_{-19}$ & 
58$^{+21}_{-18}$ & 46$^{+20}_{-18}$ & 44$^{+3}_{-2}$ \\[1.0ex]
(2)~~\texttt{powerlaw}$_2$ & & & & & \\[0.5ex]
~~~~~~~~~~$\Gamma$ & 2.0(t) & 2.0(t) & 2.0(t) & $-$ & 2.10(t) \\[0.5ex]
~~~~~~~~~~$A$ (10$^{-4}$\,cm$^{-2}$\,s$^{-1}$\,keV$^{-1}$) & 1.01$\pm$0.06 & 0.71$\pm$0.08 &
0.68$^{+0.09}_{-0.08}$ & $-$ & 0.54$^{+0.06}_{-0.05}$ \\[1.0ex]
(3)~~\texttt{pexrav} & & & & & \\[0.5ex]
~~~~~~~~~~$\Gamma$ & 2.0(t) & 2.0(t) & 2.0(t) & 2.0(t) & 2.10(t) \\[0.5ex]
~~~~~~~~~~$R$ & 0.50$^{+0.20}_{-0.13}$ & 0.74$^{+0.38}_{-0.21}$ & 0.78$^{+0.41}_{-0.22}$ & 1.24$^{+0.88}_{-0.40}$ & 
1.01$^{+0.16}_{-0.13}$ \\[1.0ex]
(4)~~\texttt{apec}$_1$ & & & & & \\[0.5ex]
~~~~~~~~~~$N_\rmn{H}$ (10$^{22}$\,cm$^{-2}$) & $-$ & $-$ & $-$ & $< 0.08$ & $-$ \\[0.5ex]
~~~~~~~~~~$kT$ (keV) & $-$ & 0.94$^{+0.07}_{-0.09}$ & 0.96$^{+0.07}_{-0.08}$ & 1.20$^{+0.10}_{-0.11}$ & 
0.99$^{+0.08}_{-0.09}$ \\[0.5ex]
~~~~~~~~~~$Z_\rmn{O}$ ($Z_{\odot}$) & $-$ & 1.0(f) & 1.0(f) & 0.03$^{+0.04}_{-0.02}$ & 1.0(f) \\[0.5ex]
~~~~~~~~~~$Z_\rmn{Ne}$ ($Z_{\odot}$) & $-$ & 1.0(f) & 1.0(f) & 0.17$^{+0.33}_{-0.13}$ & 1.0(f) \\[0.5ex]
~~~~~~~~~~$Z_\rmn{Mg}$ ($Z_{\odot}$) & $-$ & 1.0(f) & 1.0(f) & 0.36$^{+0.22}_{-0.17}$ & 1.0(f) \\[0.5ex]
~~~~~~~~~~$Z_\rmn{Si}$ ($Z_{\odot}$) & $-$ & 1.0(f) & 1.0(f) & 0.49$^{+0.18}_{-0.15}$ & 1.0(f) \\[0.5ex]
~~~~~~~~~~$Z_\rmn{Fe}$ ($Z_{\odot}$) & $-$ & 1.0(f) & 1.0(f) & 0.05$^{+0.03}_{-0.05}$ & 1.0(f) \\[0.5ex]
~~~~~~~~~~EM (10$^{63}$\,cm$^{-3}$) & $-$ & 0.25$\pm$0.06 & 0.25$^{+0.06}_{-0.05}$ & 
1.86$^{+0.23}_{-0.21}$ & 0.18$^{+0.05}_{-0.04}$ \\[1.0ex]
(5)~~\texttt{apec}$_2$ & & & & & \\[0.5ex]
~~~~~~~~~~$N_\rmn{H}$ (10$^{22}$\,cm$^{-2}$) & $-$ & $-$ & $-$ & $< 0.35$ & $-$ \\[0.5ex]
~~~~~~~~~~$kT$ (keV) & $-$ & $-$ & 0.14$^{+0.04}_{-0.03}$ & 0.15$^{+0.10}_{-0.05}$ & $-$ \\[0.5ex]
~~~~~~~~~~EM (10$^{63}$\,cm$^{-3}$) & $-$ & $-$ & 0.55$^{+0.46}_{-0.32}$ & $> 5.7^a$ & $-$ \\[1.0ex]
Fit statistic $C/\nu$ & 819/489 & 730/487 & 706/485 & 671/479 & 442/434 \\
\hline
\end{tabular}
\flushleft{Notes: $^a$Due to the fact that the peak temperature of this component falls 
outside the fitting range, the emission measure is largely degenerate with the column density, 
so that no sensible upper limit can be obtained.}
\end{table*}

\begin{figure}
\includegraphics[width=8.5cm]{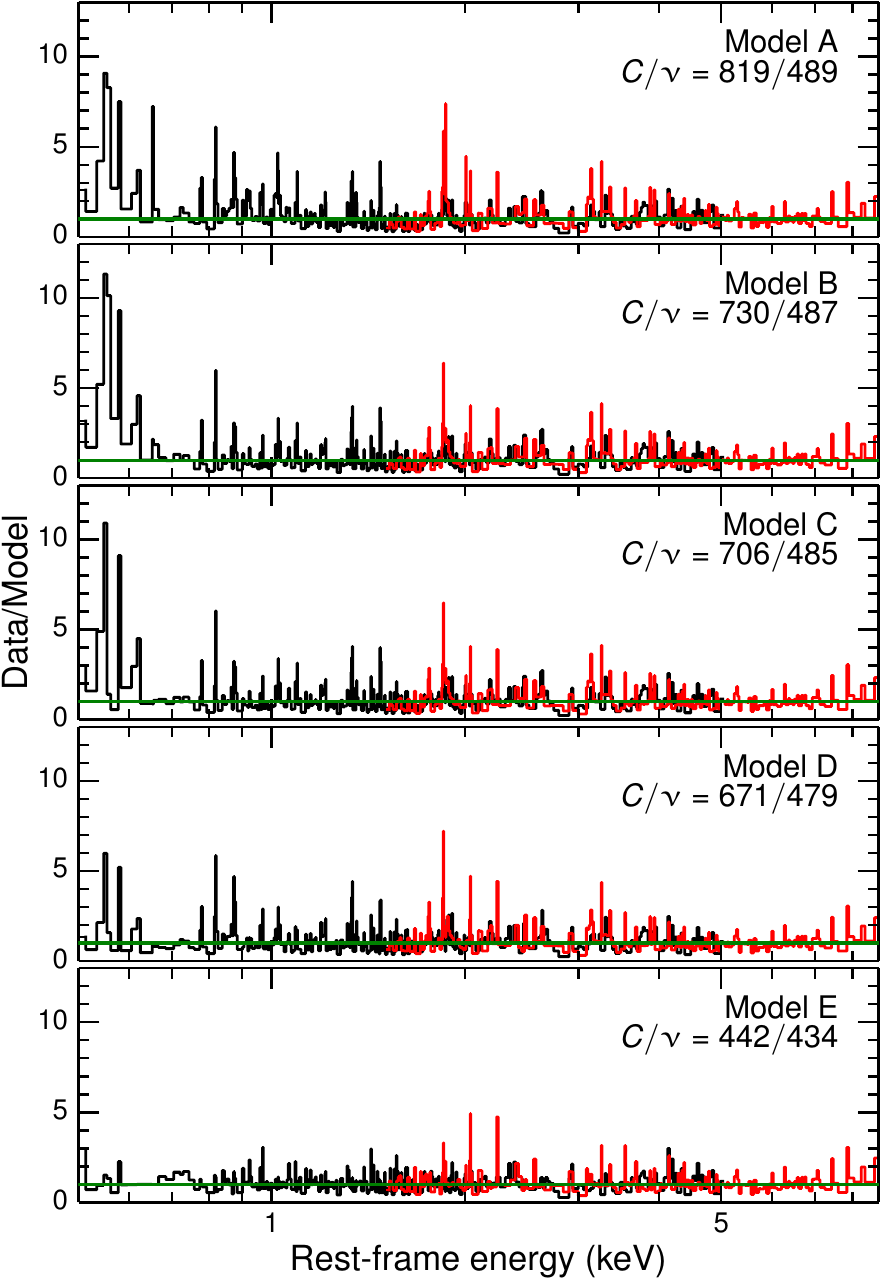}
\caption{Fit residuals for the different models A--E (error bars are omitted for clarity). 
Once the AGN-related continuum components (i.e. transmitted, reflected and scattered) and 
the 6.4-keV iron K line are accounted for, a wealth of narrow emission features is left 
below $\sim$4 keV (A). The addition of a thermal component from collisionally ionized gas 
returns an appreciable improvement around 1 keV (B), but any further thermal contribution 
is not strongly required (C,\,D). An adequate fit is obtained only after including 
individual Gaussian profiles to describe the 22 remaining emission lines detected with 
a statistical significance of at least 95 per cent (E).} 
\label{mr}
\end{figure}

\begin{table*}
\caption{Statistically significant emission lines in the \textit{Chandra} HETG spectra, after 
accounting for the collisional component. The most plausible identifications are also listed. 
(f), (i), and (r) refer to forbidden, intercombination and resonance transitions for He-like 
species. (u) stands for unresolved ($\sigma = 2$ eV is then assumed). Note that the equivalent 
widths (EW) of broad features should not be taken at face value, since in most cases these are 
blends of two (or more) narrow lines (see the text for a thorough discussion).}
\label{t2}
\begin{tabular}{r@{\hspace{5pt}}c@{\hspace{20pt}}ccc@{\hspace{20pt}}c@{\hspace{20pt}}c@{\hspace{10pt}}c}
\hline
\multicolumn{2}{c}{$E_\rmn{obs}$} & $\sigma$ & Flux & EW & 
Transition & $E_\rmn{lab}$ & $\Delta C/\Delta\nu$ \\
\multicolumn{2}{c}{(eV)} & (eV) & (10$^{-6}$\,s$^{-1}$\,cm$^{-2}$) & (eV) & & (eV) & \\
\hline
1.  & 555$^{+9}_{-8}$    & 21$^{+13}_{-7}$   & 93$^{+37}_{-32}$    & 410$^{+270}_{-220}$ & 
O\,\textsc{v} K$\alpha$                 & 554$^a$  & 54/3 \\[-0.5ex]
    &                    &                   &                     & & 
O\,\textsc{vii} He$\alpha$\,(f)         & 561$^b$  &      \\[1.0ex]
2.  & 618$^{+3}_{-2}$    & 2(u)             & 11$^{+11}_{-7}$     & 39$^{+39}_{-33}$ & 
O\,\textsc{v} K$\beta$                  & 621$^a$  & 10/2 \\[1.0ex]
3.  & 655$^{+2}_{-3}$    & 2(u)             & 7.0$^{+7.5}_{-5.3}$ & 37$\pm$34 & 
O\,\textsc{viii} Ly$\alpha$             & 654$^b$  &  6/2 \\[1.0ex]
4.  & 778$^{+3}_{-2}$    & 2(u)             & 2.9$^{+2.8}_{-2.0}$ & 18$\pm$15 & 
O\,\textsc{viii} Ly$\beta$              & 775$^b$  &  9/2 \\[-0.5ex]
    &                    &                   &                     & & 
Fe\,\textsc{xviii} $3s \rightarrow 2p$  & 775$^c$  &      \\[1.0ex]
5.  & 819$^{+1}_{-2}$    & 2(u)             & 3.7$^{+2.8}_{-2.1}$ & 22$\pm$18 & 
O\,\textsc{viii} Ly$\gamma$             & 817$^b$  & 15/2 \\[-0.5ex]
    &                    &                   &                     & & 
Fe\,\textsc{xix} $3s \rightarrow 2p$    & 822$^c$  &      \\[1.0ex]
6.  & 874$^{+6}_{-4}$    & 6$^{+9}_{-3}$     & 4.2$^{+2.8}_{-2.3}$ & 26$^{+23}_{-18}$ & 
O\,\textsc{viii} RRC                    & 871$^b$  & 12/3 \\[0.5ex]
    &                    &                   &                     & & 
Fe\,\textsc{xviii} $3d \rightarrow 2p$    & 873$^c$  &      \\[1.0ex]
7.  & 1024$\pm$4         & 6$^{+7}_{-3}$     & 2.9$^{+1.8}_{-1.5}$ & 24$^{+22}_{-17}$ & 
Ne\,\textsc{x} Ly$\alpha$               & 1022$^b$   & 14/3 \\[1.0ex]
8.  & 1096$^{+2}_{-6}$    & 2(u)            & 0.9$^{+1.0}_{-0.7}$ & 11$^{+10}_{-8}$ & 
Ne\,\textsc{ix} He$\beta$\,(r)          & 1074$^b$   &  6/2 \\[1.0ex]
9.  & 1205$\pm$6         & 8$^{+7}_{-4}$     & 1.4$^{+1.0}_{-0.8}$ & 24$\pm$17 & 
Ne\,\textsc{x} Ly$\beta$                & 1211$^b$ & 11/3 \\[1.0ex]
10. & 1334$^{+1}_{-2}$   & 2(u)            & 1.4$^{+0.9}_{-0.6}$ & 26$\pm$15 & 
Mg\,\textsc{xi} He$\alpha$\,(f)         & 1331$^b$ & 24/2 \\[1.0ex]
11. & 1356$\pm$3         & 2(u)            & 0.6$^{+0.6}_{-0.5}$ & 10$^{+12}_{-9}$ & 
Mg\,\textsc{xi} He$\alpha$\,(r)         & 1352$^b$ &  6/2 \\[1.0ex]
12. & 1476$\pm$1         & 2(u)            & 1.2$^{+0.5}_{-0.4}$ & 29$^{+17}_{-14}$ & 
Mg\,\textsc{xii} Ly$\alpha$             & 1473$^b$ & 42/2 \\[1.0ex]
13. & 1756$^{+7}_{-6}$   & 2(u)            & 0.4$\pm$2           & 10$^{+10}_{-7}$ & 
Mg\,\textsc{xii} Ly$\beta$              & 1745$^b$ &  9/2 \\[1.0ex]
14. & 1864$^{+8}_{-7}$   & 24$^{+9}_{-6}$   & 2.8$\pm$0.7         & 96$^{+67}_{-36}$ & 
Si\,\textsc{xiii} He$\alpha$\,(r)       & 1865$^b$ & 62/3 \\[1.0ex]
15. & 2006$^{+2}_{-3}$   & 2(u)            & 0.5$\pm$0.3         & 19$^{+18}_{-15}$ & 
Si\,\textsc{xiv} Ly$\alpha$             & 2006$^b$ & 10/2 \\[1.0ex]
16. & 2187$^{+7}_{-6}$   & 2(u)            & 0.9$^{+0.5}_{-0.6}$ & 40$^{+49}_{-31}$ & 
Si\,\textsc{xiii} He$\beta$\,(r)        & 2183$^b$ &  9/2 \\[1.0ex]
17. & 2475$^{+8}_{-5}$   & 2(u)            & 0.9$^{+0.7}_{-0.5}$ & 41$^{+53}_{-34}$ & 
S\,\textsc{xv} He$\alpha$\,(r)          & 2461$^b$ & 11/2 \\[1.0ex]
18. & 2630$\pm$20        & 27$^{+26}_{-19}$ & 2.0$^{+1.1}_{-1.0}$ & 100$^{+150}_{-80}$ & 
S\,\textsc{xvi} Ly$\alpha$              & 2623$^b$ & 15/3 \\[1.0ex]
19. & 3138$^{+13}_{-11}$ & 19$^{+14}_{-10}$ & 1.7$^{+0.8}_{-0.7}$ & 83$^{+110}_{-42}$ & 
Ar\,\textsc{xvii} He$\alpha$\,(r)         & 3140$^b$ & 23/3 \\[1.0ex]
20. & 3310$^{+34}_{-31}$ & 54$^{+25}_{-14}$ & 2.0$^{+1.1}_{-0.9}$ & 105$^{+205}_{-75}$ & 
Ar\,\textsc{xviii} Ly$\alpha$           & 3323$^b$ & 15/2 \\[1.0ex]
21. & 3873$^{+28}_{-30}$ & 83$^{+67}_{-22}$ & 2.8$^{+0.6}_{-0.9}$ & 160$^{+540}_{-140}$ & 
Ca\,\textsc{xix} He$\alpha$\,(f)        & 3861$^b$ & 17/3 \\[-0.5ex]
    &                    &                       &                     &  &
Ca\,\textsc{xix} He$\alpha$\,(i)        & 3883$^b$ &      \\[-0.5ex]
    &                    &                       &                     &  &
Ca\,\textsc{xix} He$\alpha$\,(r)        & 3902$^b$ &      \\[1.0ex]
22. & 6404$^{+13}_{-14}$ & 24$^{+17}_{-12}$ & 7.0$^{+2.8}_{-2.4}$ & 205$^{+80}_{-70}$ &
Fe\,\textsc{i} K$\alpha$                & 6403$^b$ & 38/3 \\
\hline
\multicolumn{8}{l}{References: $^a$Kaastra et al. (2004); $^b$NIST (\url{http://physics.nist.gov/asd}); 
$^c$CHIANTI (Landi et al. 2012).}
\end{tabular}
\end{table*}

Model A already provides a good representation of the observed continuum, while the intrinsic 
power law is largely unconstrained at this stage, also because of the decreasing data quality 
at higher energies; the photon index and absorbing column were then frozen to the reasonable 
values of $\Gamma = 2$ and $N_\rmn{H} = 2 \times 10^{24}$ cm$^{-2}$. The overall statistics 
of $C/\nu = 819/489$ are still rather poor without accounting for the emission lines (Fig. 
\ref{mr}). To check whether some of these can be ascribed to optically-thin, collisionally 
ionized plasma, which is indeed a major ingredient of the \textit{XMM--Newton}/RGS spectrum 
(Guainazzi et al. 2009), we added an \texttt{apec} thermal component (Smith et al. 2001) with 
solar abundances. In this scenario, the ionization is driven by the impact of free electrons 
with a temperature comparable with the energy of the observed lines. The new fit (model B) 
brings a substantial improvement, with $\Delta C = -89$ for the loss of two degrees of freedom 
only, taken by the plasma temperature $kT \simeq 0.95$ keV and emission measure.\footnote{The 
emission measure (in cm$^{-3}$) is defined as $\rmn{EM} = \int{n_e n_\rmn{H} \rmn{d}V}$, where 
$n_e$ and $n_\rmn{H}$ are the electron and hydrogen densities.} The most affected is the 
$\sim$0.6--1.2 keV (10--20\,\AA) range, where there are hints of iron L-shell emission, mainly 
due to $3d \rightarrow 2p$ transitions from Fe\,\textsc{xviii}--\textsc{xxiii}. This hot gas 
can be naturally identified with the residual contamination from the circumnuclear star formation 
(Fig.~\ref{im}). The related continuum is much fainter than in the large-aperture RGS data 
(see also Fig.~\ref{xs}), and the contribution to the most prominent features is no larger 
than $\sim$30--40 per cent in the case of Ly$\alpha$ lines from O\,\textsc{viii}, 
Ne\,\textsc{x}, and Mg\,\textsc{xii}. We therefore introduced a second \texttt{apec} 
component of different temperature, but this is remarkably cold ($kT \simeq 0.15$ keV; 
Table~\ref{t1}) and simply imparts minor changes to the noisy lower end of the MEG 
spectrum ($C/\nu = 706/485$; model C). Apparent structures compatible with the He-like 
triplets from Mg\,\textsc{xi}, Si\,\textsc{xiii} and Ar\,\textsc{xvii} are left unexplained, 
but no other zone of thermal emission is statistically required. 

As an ultimate attempt to explore the significance of collisionally ionized plasma, we 
also allowed for variable abundances (tied between the two components) for all the key 
$\alpha$-elements (O, Ne, Mg and Si) plus Fe, and for the possible column density 
enshrouding each star-forming region. The scattered AGN continuum was neglected to avoid 
any degeneracies (model D). The same set of assumptions was successfully applied by Guainazzi 
et al. (2009) in reproducing both the RGS and the previous \textit{Chandra}/ACIS spectra of 
NGC\,1365. With respect to their description, in the HETG data the separation between the 
plasma temperatures is much more pronounced, $\sim$0.15--1.2 keV against $\sim$0.3--0.7 
keV, while absolute and relative abundances are rather similar (once the conversion 
between the adopted solar standards is applied). In any case, the global statistics only 
undergo a scarce decrease ($\Delta C = -35$ for $\Delta \nu = -6$), and the fit stays far 
from being satisfactory ($C/\nu = 671/479$). The full details are listed in Table~\ref{t1}. 
Equivalent results were achieved by letting vary a common $\alpha$-element metallicity 
over the two regions ($Z_\alpha \sim 0.5$ and 0.03). 

Incidentally, we note that model D actually delivers a fairly good fit ($C/\nu = 513/488$) 
of the zeroth-order spectrum at CCD resolution,\footnote{A merged, low-resolution spectrum 
was extracted from the undispersed images, but is not discussed any further in this work.} 
although with slightly adjusted values ($kT \sim 0.9$ and 0.3 keV, $N_\rmn{H} \sim 1.5$ and 
$< 0.2 \times 10^{22}$ cm$^{-2}$; abundances are poorly constrained but are consistent with 
being solar). We conclude that the copious line emission revealed by the \textit{Chandra}/HETG 
observations is dominated by photoionized gas directly exposed to the radiation field of the 
AGN, which is clearly brought out for the first time.

\subsection{Emission lines}

While several star-forming regions are present within in the inner kpc, their weak soft 
X-ray emission precludes an accurate determination of the full range of properties (e.g. 
temperatures and abundances) of the hot diffuse gas. We therefore built on model B, with 
a single thermal component of solar metallicity besides the transmitted, reflected and 
scattered AGN continua. A Gaussian profile was then added for any residual emission line, 
provided that each inclusion resulted in a $\Delta C = -6.0$ or $-7.8$, corresponding to 
the 95 per cent confidence level for either two or three parameters of interest (model E). 
Following this criterion, which is less stringent than usual but is more appropriate to 
the signal to noise of the current data, 22 lines are detected (Table~\ref{t2}). At a 
first glance, based on the most likely identifications, all the main transitions of 
$\alpha$-elements from O to Ca appear to be present. Most lines are narrow, and some 
of the broad ones are possible blends. Whenever the width turned out to be consistent 
with zero at the 90 per cent level, $\sigma = 2$ eV was assumed. This does not automatically 
imply that a line is unresolved, due to its low number of counts and significance. We will 
address this point again in the next Section. 

Interestingly, all the continuum components were left free to vary and returned very sensible 
values (Table~\ref{t1}), leading to an excellent fit with $C/\nu = 442/434$. The photon index 
of the intrinsic AGN power-law emission is $\Gamma \simeq 2.10$, and the gas column density is 
$N_\rmn{H} \simeq 1.4 \times 10^{24}$ cm$^{-2}$; 
the absorption-corrected 2--10 keV flux is of the order of $\sim$10$^{-11}$ erg s$^{-1}$ 
cm$^{-2}$, in keeping with the historical behaviour of NGC\,1365. The strength of the reflection 
component is ordinary as well, $R \simeq 1$, and only $\sim$1 per cent of the primary AGN 
continuum is scattered back into our line of sight. As for the collisionally ionized plasma, 
its characteristic temperature converges to $kT \simeq 1$ keV, contributing to most of the 
Fe L-shell forest and to some extent also to the Ly$\alpha$ lines from the main $\alpha$-elements 
(see above). With this respect, the spectral parameters reported in Table~\ref{t2} only refer 
to the photoionized component. Below we discuss in greater detail the properties of the various 
lines and their implications on the physical conditions of the emitting gas. 
 
\section{Discussion}

When seen at high resolution, the soft X-ray spectra of obscured active galaxies 
display a wealth of narrow emission lines with little or no continuum, arising from 
low-temperature ($T < 10^5$ K, or a few eV) gas photoionized and photoexcited by the 
primary nuclear radiation (e.g. Sako et al. 2000; Kinkhabwala et al. 2002; Schurch 
et al. 2004). \textit{Chandra} images have also revealed the substantial spatial 
extension of the soft X-ray emission, and its striking correlation with the size 
and morphology of the Narrow Line Region (NLR) as mapped by \textit{HST} in the 
[O\,\textsc{iii}] $\lambda$5007 line (Ogle et al. 2000; Bianchi, Guainazzi \& Chiaberge 
2006). This suggests that the same gas shines in both [O\,\textsc{iii}] and soft 
X-rays. In general, the latter can also be a signature of mechanical and shock heating 
in star-forming regions, yet the contribution from collisionally ionized plasma is 
usually minor (Brinkman et al. 2002). Although detailed spectral diagnostics are 
available for the handful of brightest sources only (e.g. Kallman et al. 2014), 
this trend seems to apply to the entire population of local type 1.5--2 Seyfert 
galaxies, with just a few exceptions (Guainazzi \& Bianchi 2007). NGC\,1365 stands 
out among the peculiar objects by virtue of the foremost influence of thermal plasma 
on its RGS spectrum. According to Guainazzi et al. (2009), evidence for photoionized 
matter solely relies on excess emission in the forbidden (f) and intercombination (i) 
components of O\,\textsc{vii} He$\alpha$, in N\,\textsc{vi} He$\alpha$\,(f), and in the 
tentative O\,\textsc{vii} and O\,\textsc{viii} radiative recombination continua (RRC). 

The new \textit{Chandra}/HETG observations take advantage of a much better angular 
resolution, and allowed us to ignore the bulk of the circumnuclear starburst. The 
resultant spectra cannot be explained through a multi-temperature gas in collisional 
equilibrium, whose importance is conversely highly diminished. The emergence of the 
diluted emission from a photoionized region closer to the hard X-ray source tends to 
reconcile the nature of NGC\,1365 with the average properties of the local obscured 
AGN. In order to move towards a self-consistent description of the whole sequence of 
lines in Table~\ref{t2}, we removed all the Gaussian profiles from model E (except 
for Fe\,\textsc{i} K$\alpha$) and generated several sets of photoionization grids 
with \textsc{xstar} (Kallman \& Bautista 2001). We adopted an optical to X-ray spectral 
energy distribution similar to the one used in Guainazzi et al. (2009), based on an 
$\alpha_\rmn{ox} = 1.5$ energy index,\footnote{The spectral index $\alpha_\rmn{ox} = 
-0.384 \log (L_\rmn{2\,keV}/L_\rmn{2500\,\text{\AA}})$ is the usual indicator of the 
optical to X-ray luminosity ratio in AGN.} and on a 1--1000 Rydberg (13.6 eV to 13.6 
keV) luminosity of $\sim$10$^{44}$ erg s$^{-1}$. For each turbulence velocity broadening 
of 500, 1000, 1500 and 2000 km s$^{-1}$ (1$\sigma$), the grids cover five orders of 
magnitude in both column density and ionization parameter,\footnote{The ionization 
parameter is defined as $\xi = L_\rmn{ion}/nr^2$, where $n$ is the gas electron density 
and $r$ is its distance from the source with 1--1000 Ry luminosity $L_\rmn{ion}$.} with 
$5 \times 10^{19} < N_\rmn{H}/\rmn{cm^{-2}} < 5 \times 10^{24}$ and 
$0 < \log\,(\xi/\rmn{erg\,cm\,s^{-1}}) < 5$, respectively. We also tested different 
gas densities, typical of BLR and NLR clouds, with indistinguishable results. 

The inclusion of any photoionized component was accepted as significant after a statistical 
improvement of $\Delta C = -13.3$, corresponding to the 99 per cent confidence level for 
four parameters of interest, i.e. $N_\rmn{H}$, $\xi$, normalization and outflow velocity. 
As expected, no exhaustive solution was found in this effort, which should be merely taken 
as a rudimentary proof of the existence of a fairly large range of ionization states. This 
notwithstanding, a passable fit ($C/\nu = 605/479$) was achieved with three emission phases 
with $\log\,(\xi/\rmn{erg\,cm\,s^{-1}}) > 4.1$, $\sim$3.1$\pm$0.3, and $< 0.6$. The other 
gas properties are not well constrained (especially the column densities), so we refrain 
from discussing their face values. Anyway, there is evidence for motions along the line 
of sight: two \textsc{xstar} components have a modest outflow velocity of $v_\rmn{out} = 
850^{+620}_{-330}$ km s$^{-1}$, while the third one prefers an inflow at $\sim$3500 km 
s$^{-1}$ to a systemic velocity ($\Delta C = 7$). This is driven by the broad complex 
centred at $E \sim 555$ eV (Table~\ref{t2}) and its identification with O\,\textsc{vii} 
He$\alpha$, which is not fully reliable (see below). In this model, once the threefold 
AGN continuum is excluded, the contribution from collisionally ionized plasma to the 
0.5--4 keV flux is about 30 per cent. Overall, the main shortcoming is that the intensity 
of many of the former Gaussian lines is slightly underpredicted, possibly due to higher 
metallicity. A 3$\times$ solar Fe abundance in the nuclear region of NGC\,1365 is indeed 
suggested by the X-ray reflection features (Risaliti et al. 2009b; Brenneman et al. 2013).

\begin{figure*}
\includegraphics[width=0.95\textwidth]{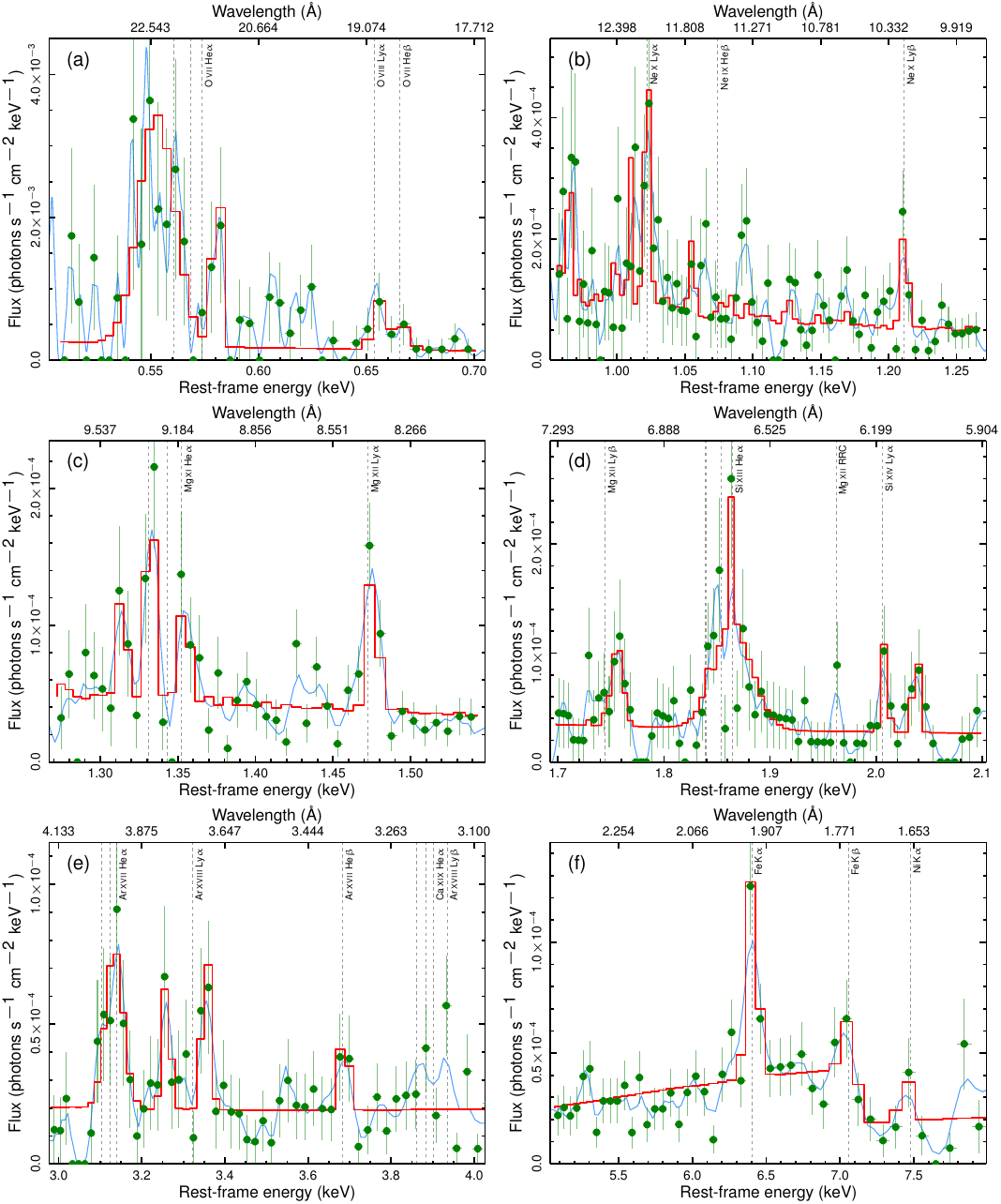}
\caption{Enlarged view of the MEG (a--c) and HEG (d--f) spectra in the main emission bands. 
The thermal and AGN continuum components were assumed from model E, while the lines were 
refitted (red curve). The data were rebinned to 1024 channels, corresponding to $\sim$2 times 
the native FWHM resolution. For the sake of a visual inspection, a 5-channel wide triangular 
smoothing has been applied to the ungrouped spectra (turquoise curve). The vertical lines mark 
the expected energy of the stronger transitions. (a) Due to the poor signal to noise, the sampling 
in the oxygen band is further degraded to a factor of 8, with a smoothing window of 13 channels. 
(b) Both Ne\,\textsc{x} Ly$\alpha$ and Ly$\beta$ are detected with a possible velocity broadening 
($\rmn{FWHM} \sim 1300$ km s$^{-1}$). The thermal component is not able to fully account for all 
the putative Fe lines. (c) Despite the triplet-like appearance of the emission complex at 
$\sim$1.31--1.36 keV, the lower-energy line is not associated with Mg\,\textsc{xi} He$\alpha$ 
and is most likely due to Fe\,\textsc{xxi} $4d \rightarrow 2p$. (d) The broad resonance line 
of Si\,\textsc{xiii} can be resolved into a blueshifted ($v_\rmn{out} \sim 1600$ km s$^{-1}$) 
triplet. (e) The double peak across Ar\,\textsc{xviii} Ly$\alpha$ would imply inflows/outflows 
at several thousands km s$^{-1}$, but the red horn is not confirmed in the MEG data. (f) Besides 
the resolved Fe K$\alpha$ feature, both Fe K$\beta$ and Ni K$\alpha$ are clearly present although 
statistically not significant due to the low effective area over this range. (See the text for 
more details).} 
\label{lz}
\end{figure*}

\subsection{Line properties}
For the sake of a qualitative assessment, we subsequently considered both the MEG and HEG 
spectra at 1024 and 2048 channels, preserving the constant resolution in wavelength space 
afforded by the gratings. This choice is not suitable for global statistical purposes, since 
no photons are collected in several bins, especially at very low or high energies. Moreover, 
because of the low number of counts, the aspect and position of some lines is contingent on 
the binning criterion adopted. While some degree of speculation is admitted in this heuristic 
approach, we always checked the raw data in wavelength units to avoid exotic interpretations, 
sticking to the most conservative explanation. None the less, in many cases this helped us 
extracting more detailed information on the properties of the emission lines, for instance 
in terms of widths, velocity shifts and relative weight of the different components in He-like 
triplets. To this aim, we also created smoothed spectral profiles by applying a triangular 
moving average to the unbinned (2048 channels) data. For each band of interest, the results 
of this procedure are shown in Fig.~\ref{lz}, where the baseline continuum is drawn from 
model E and the key features have been fitted afresh. In the following we examine every 
atomic species separately, and then try to infer a more general picture. 

\subsubsection{Oxygen}

Unfortunately, at the energies of the O\,\textsc{vii}--\textsc{viii} emission the MEG spectrum 
is extremely noisy, and perhaps also affected by some background. In this case, we had to rebin 
down to 256 channels and perform a comparably heavy smoothing (Fig.~\ref{lz}a). Even so, it is 
clear that the broad feature at $\sim$555 eV in model E (which could be blindly associated with 
O\,\textsc{v} K$\alpha$ from its centroid) presents several substructures, and that the line at 
$E \sim 618$ eV in Table~\ref{t2} (in formal agreement with O\,\textsc{v} K$\beta$) is actually 
a blend (possibly including N\,\textsc{vii} Ly$\gamma$). We attempted at fitting the former 
complex with a self-consistent O\,\textsc{vii} He$\alpha$ triplet, fixing the energy of the three 
components and allowing for a common shift. As anticipated above, this selects a redshift with 
velocity of $\sim$5500 km s$^{-1}$, which would be loosely supported by the identification of 
the line at $E \simeq 581$ eV (falling just below our detection threshold) with N\,\textsc{vii} 
Ly$\beta$ ($E \simeq 593$ eV at rest). Bearing in mind some claims in this sense (like for 
Mrk\,110; Boller, Balestra \& Kollatschny 2007), such a fast inflow is not regarded as genuine 
in this context, and the $\sim$555-eV blend remains ambiguous. We note that one of the spikes in 
the smoothed spectrum has the right energy for the forbidden component of the oxygen He-like 
triplet, but other species might be involved (for example, N\,\textsc{vi} RRC is found at 552 eV). 
The statistics are seriously deficient, but the feature at $E \simeq 581$ eV can be tentatively 
ascribed to outflowing O\,\textsc{vii}. Also the O\,\textsc{viii} Ly$\alpha$ line suggests a 
$v_\rmn{out} \sim 500\pm200$ km s$^{-1}$, and the higher order Ly$\beta$, Ly$\gamma$ and RRC 
transitions of this ion are consistent with this value,\footnote{In the O\,\textsc{viii} energy 
range, the MEG absolute wavelength accuracy is equivalent to $\sim$200 km s$^{-1}$.} even if 
suffering from contamination by the Fe L-shell array.

\subsubsection{Neon}

There is very little evidence for Ne\,\textsc{ix} emission, once the almost coincident 
$3d \rightarrow 2p$ lines from Fe\,\textsc{xix} are attributed to collisionally ionized 
gas. Ne\,\textsc{x} Ly$\alpha$ and Ly$\beta$, instead, are both obvious and definitely 
interesting. They appear to be resolved ($\sigma = 1.9^{+0.7}_{-0.5}$ eV for $\Delta C 
= 1$) and minimally blueshifted ($v_\rmn{out} < 400$ km s$^{-1}$; Table~\ref{t3}). 
The intensity ratio Ly$\beta$/Ly$\alpha$ is quite large, $\sim$0.6$\pm$0.3. This is a hint 
of photoexcitation, which boosts the higher order lines through radiative decay (e.g. 
Kinkhabwala et al. 2002). Also Fe L-shell resonances would be enhanced (Sako et al. 2000), 
and indeed the thermal component fails to reproduce their emission at $\sim$1.06--1.18 keV 
(Fig.~\ref{lz}b). Notably, the strongest $3d \rightarrow 2p$ Fe lines, including 
Fe\,\textsc{xx} (0.956 keV) and Fe\,\textsc{xxi} (1.009 keV), accept some blueshift at 
$v_\rmn{out} \sim 900$ km s$^{-1}$. This is probably too high (e.g. Veilleux, Cecil \& 
Bland-Hawthorn 2005) for any starburst superwind in NGC\,1365, favouring an origin from 
matter photoionized and photoexcited by the AGN rather than shock-heated much farther out. 

\subsubsection{Magnesium}

Three lines are seen in the Mg\,\textsc{xi} He$\alpha$ band (Fig.~\ref{lz}c), but the 
low-energy one is actually due to Fe\,\textsc{xxi} $4d \rightarrow 2p$, again more 
prominent than in a pure thermal scenario. The remaining pair are presumably the 
forbidden and recombination components of the triplet. Together with the Mg\,\textsc{xii} 
Ly$\alpha$ line, they share $\sigma = 1.6 \pm 0.6$ eV ($\Delta C = 1$) and $v_\rmn{out} 
\sim 700\pm100$ km s$^{-1}$ (Table~\ref{t3}). The only other feature from these ions is 
the possible Mg\,\textsc{xii} RRC, while the puzzling excess around the energy of the 
Ly$\beta$ is formally incompatible with the shift of the parent series, and is eventually 
unknown (Fig.~\ref{lz}d). 

\begin{figure}
\includegraphics[width=8.5cm]{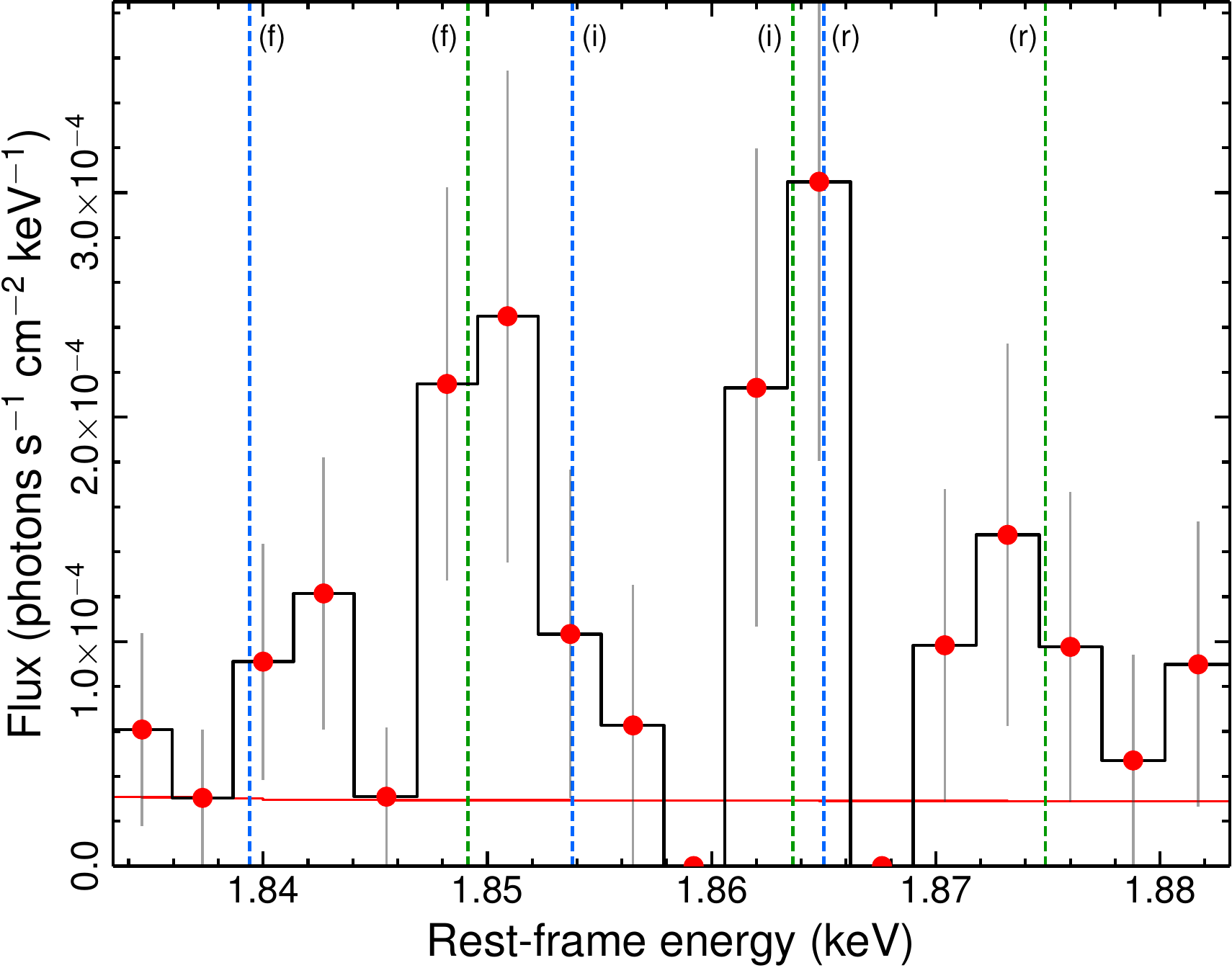}
\caption{Si\,\textsc{xiii} He$\alpha$ triplet in the HEG spectrum at 2048 channels 
($\Delta\lambda \simeq 10~\rmn{m\AA}$, consistent with the FWHM resolution). The 
blue and green dashed lines indicate the energy of the forbidden, intercombination, 
and resonance transitions for $v_\rmn{out} = 0$ and 1600 km s$^{-1}$, respectively. 
Both velocity components seem to be present within the energy/wavelength calibration 
accuracy, the former possibly associated with the starburst and the latter to the 
AGN. The continuum is drawn in red.} 
\label{tr}
\end{figure}

\subsubsection{Silicon} 

A broad line is detected in model E at the energy of Si\,\textsc{xiii} He$\alpha$, centred 
around the resonance transition. Its width in the HEG spectrum at 1024 channels is $\sigma \sim 
18$ eV, and a narrow core is noticeable (Fig.~\ref{lz}d). While the FWHM of $\sim$7000 km 
s$^{-1}$ would dictate a BLR origin, we tried to disentangle the triplet components. The best 
solution is found for unresolved profiles at $v_\rmn{out} \sim 1600\pm100$ km s$^{-1}$, with 
the forbidden and resonance lines shifted to $\sim$1.850 and 1.875 keV (Fig.~\ref{tr}). This 
outflow velocity is not confirmed by Si\,\textsc{xiv} Ly$\alpha$, which lies at the expected 
rest-frame energy. Moreover, after the joint $\sim$10-eV shift, the intercombination line 
becomes the stronger one at $\sim$1.865 keV, which is inconsistent with the weights of 
Mg\,\textsc{xi} and practically unphysical. On the other hand, its intensity is naturally 
explained by a contribution from resonant emission at rest, supported by the faint forbidden 
companion and associated with either the AGN or the starburst. Two velocity components are 
tentatively distinguished then. Only the properties of the blueshifted triplet are listed 
in Table~\ref{t3}. Finally, there is an unidentified secondary peak at $E \sim 2.039$ keV. 
The closer transition is Al\,\textsc{xiii} Ly$\beta$ ($E \simeq 2.048$ keV), but any Ly$\alpha$ 
counterpart would be much weaker. If due to H-like silicon, the sizable shift calls for 
$v_\rmn{out} \sim 4700$ km s$^{-1}$. 

\subsubsection{Argon} 

While S\,\textsc{xv}--\textsc{xvi} is recognized but very faint, some basic constraints 
can be put on Ar\,\textsc{xvii}--\textsc{xviii}. The broad ($\sigma \sim 17$ eV) He$\alpha$ 
line is found again at the resonance energy. The same considerations made above for the 
Si triplet could apply, but here any fine structure cannot be appreciated since the 
separation of the three components is commensurate with the FWHM. The same holds for 
He$\beta$. Another double-peaked feature is visible across the Ar\,\textsc{xviii} 
Ly$\alpha$ transition (Fig.~\ref{lz}e). The red horn is dubious, since the HEG and MEG 
data are in conflict (the MEG resolution, however, is lower by a factor of 2). The blue 
one would correspond to $v_\rmn{out} \sim 2700$ km s$^{-1}$. In any case, we can rule out 
a contribution of the Ar\,\textsc{xviii} Ly$\beta$ line (3.936 keV at rest) to the blend 
near $\sim$3.90 keV, which is the alleged combination of forbidden and resonant Ca\,\textsc{xix} 
He$\alpha$ with modest blueshift. 

\begin{table}
\caption{Summary of the best constrained lines. No errors are given for tied (or frozen) 
parameters. Uncertainties are purely statistical at the 68 per cent confidence level 
($\Delta C = 1$).}
\label{t3}
\begin{tabular}{l@{\hspace{20pt}}c@{\hspace{10pt}}c@{\hspace{10pt}}c@{\hspace{10pt}}c}
\hline
\multirow{2}{*}{Line} & $E_\rmn{lab}$ & $\sigma$ & FWHM & $v_\rmn{out}$ \\
 & (keV) & (eV) & (km s$^{-1}$) & (km s$^{-1}$) \\
\hline
Ne\,\textsc{x} Ly$\alpha$ & 1.022 & 1.9$^{+0.7}_{-0.5}$ & 1280$^{+480}_{-510}$ & $< 400$ \\
Ne\,\textsc{x} Ly$\beta$ & 1.211 & 2.1 & 1280 & $< 400$ \\
Mg\,\textsc{xi} He$\alpha$\,(f) & 1.331 & 1.6$\pm$0.6 & 840$^{+300}_{-320}$ & 710$\pm$110 \\
Mg\,\textsc{xi} He$\alpha$\,(r) & 1.352 & 1.6 & 840 & 710 \\
Mg\,\textsc{xii} Ly$\alpha$ & 1.473 & 1.8 & 840 & 710 \\
Si\,\textsc{xiii} He$\alpha$\,(f) & 1.839 & $< 1.6$ & $< 600$ & 1590$^{+80}_{-60}$ \\
Si\,\textsc{xiii} He$\alpha$\,(r) & 1.865 & $< 1.6$ & $< 600$ & 1590 \\
Si\,\textsc{xiv} Ly$\alpha$ & 2.006 & $< 1.7$ & $< 600$ & $< 300$ \\
\hline
\end{tabular}
\end{table}

\subsubsection{Iron}

\begin{figure}
\includegraphics[width=8.5cm]{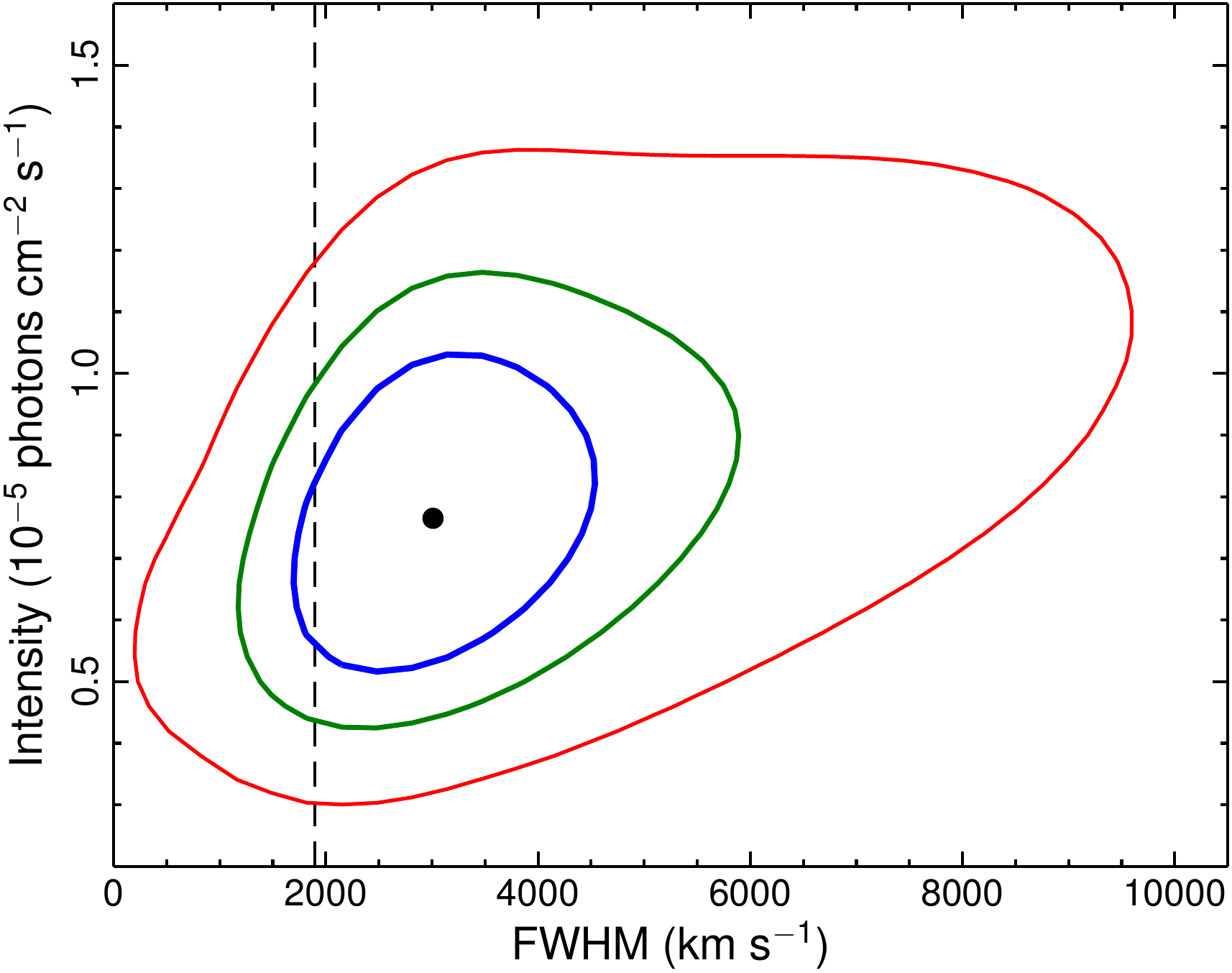}
\caption{Confidence contours at the nominal 68, 90 and 99 per cent level for the Fe K$\alpha$ 
intensity versus velocity width, obtained from the unbinned HEG spectrum with all the AGN 
continuum parameters free to vary. The vertical dashed line indicates the FWHM of the broad 
H$\beta$ component in the optical, suggesting a common origin from BLR clouds.}
\label{ka}
\end{figure}

Above 4 keV, only the fluorescence K$\alpha$ line from neutral iron at 
$E = 6.404^{+0.012}_{-0.015}$ is statistically significant, although the K$\beta$ is clear 
as well, and there is a trace of Ni K$\alpha$ (Fig.~\ref{lz}f). All of these are signatures 
of reflection of the AGN radiadion in cold, optically thick matter, arguably located at the 
torus scale in the original unification scheme. We can assess the distance of this gas from 
the width of the Fe K$\alpha$ line, which is nicely resolved thanks to its $\sim$35--40 counts 
($\sigma = 27^{+17}_{-12}$ eV; Fig.~\ref{ka}). We assume that the broadening is due to the 
virial motion in the emitting region, and that the velocity dispersion is related to the 
FWHM through a factor $f = \sqrt{3}/2$, which corrects for the (unknown) geometry and 
distribution of the gas (see also Netzer \& Marziani 2010). The predicted FWHM is 
$\sim$760($M_8/r_\rmn{pc}$) km s$^{-1}$, where $M_8$ is the mass of the central black hole 
in units of 10$^8 M_{\sun}$ and $r_\rmn{pc}$ is the radial distance in pc. This expression 
agrees within a factor of $\sim$2 with the case of pure Keplerian motion with reasonable orbital 
inclination ($i > 20\degr$). Far larger uncertainties are associated with the black hole mass. 
No direct measurement (e.g. from reverberation mapping) is available for NGC\,1365, so that 
the full range of $\log\,(M_\rmn{BH}/M_{\sun}) \sim 6$--8 is covered in the literature. The 
higher values are based on the correlation of $M_\rmn{BH}$ with the bulge luminosity, and are 
likely biased by the starburst contribution to the latter. The most sensible guess is then 
of a few $\times$10$^6\,M_{\sun}$ (Risaliti et al. 2009b; Davis et al. 2014). With a conservative 
$M_8 = 0.1$, and the observed K$\alpha$ FWHM of $\sim$3000$^{+1900}_{-1350}$ km s$^{-1}$, we 
obtain that $r \sim 0.002$--0.02\ pc, which is within the BLR. Indeed, the width of the 
K$\alpha$ line is consistent (at 90 per cent errors) with that of the broad nuclear component 
of the H$\beta$ in the optical, whose FWHM is 1895 km s$^{-1}$ (Schulz et al. 1999). From the 
K$\alpha$/H$\beta$ width ratio we can deduce that the relative size of the respective emission 
regions is roughly the same. This is not surprising in a source like NGC\,1365, where BLR 
clouds are responsible for variable X-ray obscuration up to the Fe-K band and beyond. 
Regardless of the precise black hole mass, in terms of the gravitational radius the distance 
of the Fe K$\alpha$ reflector perfectly agrees with the universal location of a few 
$\times$10$^4\,r_{g}$ determined by Shu, Yaqoob \& Wang (2011). 

\subsection{Covering factor}

As established from the FWHM of Fe K$\alpha$ (3000 km s$^{-1}$) and, provisionally, from 
those of Ne\,\textsc{x} ($\sim$1300 km s$^{-1}$) and Mg\,\textsc{xii} Ly$\alpha$ ($\sim$800 
km s$^{-1}$), the gas probed by these \textit{Chandra}/HETG spectra is likely located across 
the boundary between the classical BLR and NLR (see also Crenshaw \& Kraemer 2007). It is 
therefore tempting to try to evaluate its covering fraction ($f_\rmn{cov}$) and to explore 
its connection with the clumps responsible for the X-ray occultations of the source. To do 
so, we applied an \textsc{xstar} grid to the 1.3--1.5 keV band, characterized by the intense 
lines from Mg\,\textsc{xi}--\textsc{xii}. For given column density and amount of ionizing 
radiation, the measured Mg luminosity of $L_\rmn{Mg} \sim 6 \times 10^{38}$ erg s$^{-1}$ 
depends on the fraction of solid angle occupied by the gas. The predicted luminosity for 
a full covering of 4$\upi$ sr (i.e., $f_\rmn{cov} = 1$) is degenerate with the column 
density, but becomes slowly responsive to its exact value at $N_\rmn{H} > 2 \times 10^{23}$ 
cm$^{-2}$, which is typical for BLR clouds in NGC\,1365 (Risaliti et al. 2009b). In that 
range, the entailed $f_\rmn{cov}$ lies between $\sim$2--$7 \times 10^{-3}$. This is not far 
from the NLR covering factor delivered by photoionization models, which, however, is 
thought to be substantially underestimated (Netzer \& Laor 1993). Indeed, it should be 
quite similar to the BLR one ($\sim$0.2--0.4; e.g. Korista, Ferland \& Baldwin 1997), 
even if showing a stronger decreasing trend with luminosity (Stern \& Laor 2012). A 
column of $\sim$10$^{21}$ cm$^{-2}$, still plausible for the NLR, would be needed to 
match the expectations.

There are several caveats about this simple calculation, though. For instance, the nearly 
equivalent intensity of the He$\alpha$\,(f) and Ly$\alpha$ lines requires a fine tuning of 
$\log\,(\xi/\rmn{erg\,cm\,s^{-1}}) \sim 3.5$ in a single-zone model. If a wider range of 
ionization states is involved, $f_\rmn{cov}$ can be considerably higher. Allowing for two 
components with $N_\rmn{H} = 5 \times 10^{22}$ cm$^{-2}$ and $\log\,(\xi/\rmn{erg\,cm\,s^{-1}}) 
= 2.5$ and 4.5, respectively, returns a better fit and a covering fraction of $\sim$0.15. 
Many other configurations are viable, and also a mild attenuation of $L_\rmn{Mg}$ by an 
external absorption layer, maybe at host-galaxy scales, cannot be ruled out. Alternatively, 
given the narrower profile of the Mg lines compared to Fe K$\alpha$, we can speculate that 
the innermost clouds significantly reduce the ionizing flux seen by the emitting gas at 
larger distance, thus producing an apparently low $f_\rmn{cov}$ for the NLR. For 
$\log\,(\xi/\rmn{erg\,cm\,s^{-1}}) \sim 3.5$ and $n \sim 10^5$ cm$^{-3}$, Mg emission 
would occur at $r \sim 0.2$ pc, at least an order of magnitude higher than what inferred 
for Fe K$\alpha$ (remarkably proportional to the square of the FWHM ratio). This is the 
most intriguing explanation, since it is consistent with the under-luminosity in the soft 
X-ray oxygen lines noted by Guainazzi et al. (2009), despite the fact that the AGN in NGC\,1365 
is not intrinsically weak. Unfortunately, the current data quality does not warrant any 
conclusive answer.

\subsection{Temperature and density}

The presence of possible RRC from species like O\,\textsc{viii} or Mg\,\textsc{xii} 
implies that the emitting gas has a low temperature and that the ionization state is 
regulated by radiation rather than by collisional equilibrium. Radiative recombination, 
in fact, is the emission of a photon following the capture of a free electron, mainly to 
the ground state of an ion. The shape of the consequent feature is a powerful diagnostic 
of the gas temperature. In a cold plasma, only the tail of high-energy electrons overcome 
the recombination threshold, and the RRC have a narrow, line-like profile, whose width is 
$\Delta E \sim kT$ (Hatchett, Buff \& McCray 1976; Liedahl \& Paerels 1996). Both detections 
in proximity of O\,\textsc{viii} and Mg\,\textsc{xii} RRC rely on $\sim$6--7 line counts 
only in the MEG and HEG data, respectively. Hence we cannot disentangle the putative 
O\,\textsc{viii} RRC from the nearby Fe\,\textsc{xviii} L-shell transition, in particular 
if the former is moderately blueshifted. No blend with other strong lines is instead 
foreseen at the energy of Mg\,\textsc{xii} RRC, yet we remind that this feature is not 
statistically significant, so we were only able to put an upper limit of 3 eV (or $3.5 
\times 10^4$ K; $\Delta C = 1$) with a \texttt{redge} model. 

More information can be potentially derived by the He-like triplets, consisting of 
the resonance ($1s2p\ ^1\rmn{P}_1 \rightarrow 1s^2\ ^1\rmn{S}_0$), intercombination 
($1s2p\ ^3\rmn{P}_{2,1} \rightarrow 1s^2\ ^1\rmn{S}_0$), and forbidden 
($1s2s\ ^3\rmn{S}_1 \rightarrow 1s^2\ ^1\rmn{S}_0$) transitions. These provide an 
effective measure of the physical conditions of the gas (Porquet \& Dubau 2000, and 
references therein), since the intensity ratios $\mathcal{R} = \rmn{f/i}$ and 
$\mathcal{G} = \rmn{(f+i)/r}$ are very sensitive to the electron density and 
temperature, respectively. Given that O\,\textsc{vii} is too noisy, Ne\,\textsc{ix} 
is formally undetected, and S\,\textsc{xvii}, Ar\,\textsc{xvii} and Ca\,\textsc{xix} 
are too faint and/or unresolved, the only accessible species are Mg\,\textsc{xi} 
and Si\,\textsc{xiii}. Even if the triplet decompositions proposed above (which 
impose outflow velocities of $\sim$700 km s$^{-1}$ for Mg and $\sim$1600 km s$^{-1}$ 
for Si) are valid, only upper limits can be placed. The density can be anything below 
the critical value ($n \sim 10^{13}$--10$^{14}$ cm$^{-3}$) beyond which the forbidden 
components are collisionally suppressed, while temperature is constrained to be $T 
< 3 \times 10^6$ K. This is barely meaningful in the light of the appearance of 
narrow RRC features, but it is enough to exclude collisional ionization in a hot 
($kT \sim 1$ keV) plasma as the dominant mechanism in these emission-line HETG 
spectra of NGC\,1365.
  
\section{Summary and Conclusions}

In this paper we have presented the X-ray spectral analysis of the first 
\textit{Chandra}/HETG observations of the prototypical changing-look Seyfert 
galaxy NGC\,1365, which lingered in a Compton-thick state for the entire 
span of four days. The soft X-ray spectrum was thus dominated by a wealth 
of recombination lines belonging to He- and H-like ions of the most common 
light elements from oxygen to calcium, plus iron L-shell transitions. Thanks 
to the unrivaled spatial resolution offered by \textit{Chandra}, we were able 
to zoom in on the close surroundings of the AGN, and unveil a photoionized gas 
component that was largely diluted in previous observations. Indeed, the 
spectra obtained with \textit{XMM--Newton}/RGS, whose large aperture also 
encompasses all the diffuse emission from the inner 5 kpc, were heavily 
contaminated by collisionally ionized plasma, shock-heated by the fierce 
circumnuclear star-formation activity. The residual thermal contribution 
to the overall emission-line intensity now amounts to $\sim$30 per cent only. 

The emergence of a photoionized spectrum with properties analogous to those 
reported in most of the nearby obscured Seyfert galaxies partly mitigates the 
apparent anomaly of NGC\,1365. In spite of the modest statistical quality of 
the data, whereby only a handful of lines have $> 10$ counts, the picture in 
favour of cold ($kT \ll E$) gas exposed to the AGN radiation is corroborated 
by the tentative presence of narrow RRC features (O\,\textsc{viii} and 
Mg\,\textsc{xii}), and by the qualitative inspection of the He-like triplets 
(mainly from Mg and Si). Photoexcitation might be important as well, as 
indicated by the strength of some higher order transitions (e.g. Ne\,\textsc{x} 
Ly$\beta$) and of the Fe L-shell forest around $\sim$1 keV. Possible outflow 
velocities in the range $\sim$0--1600 km s$^{-1}$ are revealed, and a few lines 
(Ne\,\textsc{x}, Mg\,\textsc{xi}--\textsc{xii}) show some evidence of broadening 
($\sim$1000 km s$^{-1}$), signifying that the location of the gas is across the 
virtual BLR/NLR boundary. The K$\alpha$ fluorescence feature from neutral iron 
at $E = 6.404^{+0.012}_{-0.015}$ keV is resolved to a width of 
$\sigma = 27^{+17}_{-12}$ eV, corresponding to a FWHM of 
$\sim$3000$^{+1900}_{-1350}$ km s$^{-1}$. This is somewhat larger but fully consistent 
with the optical H$\beta$ line, supporting a comparable size of the emitting regions. 
With a preferential distance of few $\times$10$^4\,r_{g}$ from the black hole, 
it is then very likely that Fe K$\alpha$ arises from the same Compton-thick 
clouds that induce the recurrent column density jumps and the extreme variability 
seen in this source. 

Besides the main findings, there are also some unusual hints, such as of broad 
(BLR-like) profiles in resonance lines, of double peaks, and of radial velocities of 
several thousands km s$^{-1}$, even in inflow. If confirmed, these would be absolutely 
exceptional traits, but caution is mandatory given the poor statistics. As opposed to 
standard type-2 Seyferts like NGC\,1068, the erratic behaviour of NGC\,1365 below 10 
keV makes it hard to amass strictly consistent information from different epochs, so 
naively it could seem unfeasible to overcome this limitation with just a longer exposure.  
However, even in a Compton-thin case, the column density is normally of the order of 
$N_\rmn{H} \sim 10^{23}$ cm$^{-2}$, and the direct AGN continuum is almost completely 
absorbed in the soft X-rays. The ample record of past observations suggests that a nearly 
unobscured state such as the one sampled in January 2013 is a very rare occurrence for 
NGC\,1365. The photoelectric cutoff is thus expected to fall beyond the main complexes 
from O, Ne, Mg, and maybe also Si, leaving their aspect virtually constant with time 
and allowing the build-up of a nuclear emission-line template. On the other hand, the 
appearance of the transmitted AGN continuum at higher energies would enable the study 
of the high-ionization absorption features originating from the accretion-disc/BLR wind, 
from Fe\,\textsc{xxv}--\textsc{xxvi} down to Ca\,\textsc{xix}--\textsc{xx} and even 
Ar\,\textsc{xvii}--\textsc{xviii}, depending on the exact value of $N_\rmn{H}$. For these 
reasons, any prospective observation of NGC\,1365 at high spatial and spectral resolution 
would provide further invaluable insights into the properties of the sub-pc scale 
environment in active galaxies.

\section*{Acknowledgments}

We thank the anonymous referee for their useful comments. EN is supported 
by STFC under grant ST/J001384/1. JNR acknowledges the financial support 
provided through the \textit{Chandra} award GO2-13123A for this programme. 
The scientific results reported in this article are based on observations 
made by the \textit{Chandra} X-ray Observatory. This research has made use 
of software provided by the \textit{Chandra} X-ray Center (CXC) in the 
application package \textsc{ciao}. The figures were generated using 
\texttt{matplotlib} (Hunter 2007), a \textsc{python} library for publication 
of quality graphics.



\label{lastpage}


\begin{thebibliography}{}
\bibitem[\protect\citeauthoryear{Alonso-Herrero et al.}{2012}]{2012MNRAS.425..311A} 
Alonso-Herrero A., et al., 2012, MNRAS, 425, 311 
\bibitem[\protect\citeauthoryear{Antonucci}{1993}]{1993ARA&A..31..473A} 
Antonucci R., 1993, ARA\&A, 31, 473 
\bibitem[\protect\citeauthoryear{Antonucci \& Miller}{1985}]{1985ApJ...297..621A} 
Antonucci R.~R.~J., Miller J.~S., 1985, ApJ, 297, 621
\bibitem[\protect\citeauthoryear{Bianchi, Guainazzi, \& Chiaberge}{2006}]{2006A&A...448..499B} 
Bianchi S., Guainazzi M., Chiaberge M., 2006, A\&A, 448, 499 
\bibitem[\protect\citeauthoryear{Bianchi, Maiolino, \& Risaliti}{2012}]{2012AdAst2012E..17B} 
Bianchi S., Maiolino R., Risaliti G., 2012, AdAst, 2012, 782030
\bibitem[\protect\citeauthoryear{Boller, Balestra, \& Kollatschny}{2007}]{2007A&A...465...87B} 
Boller T., Balestra I., Kollatschny W., 2007, A\&A, 465, 87
\bibitem[\protect\citeauthoryear{Braito et al.}{2014}]{2014ApJ...795...87B} 
Braito V., Reeves J.~N., Gofford J., Nardini E., Porquet D., Risaliti G., 2014, ApJ, 795, 87
\bibitem[\protect\citeauthoryear{Brenneman et al.}{2013}]{2013MNRAS.429.2662B} 
Brenneman L.~W., Risaliti G., Elvis M., Nardini E., 2013, MNRAS, 429, 2662
\bibitem[\protect\citeauthoryear{Brinkman et al.}{2002}]{2002A&A...396..761B} 
Brinkman A.~C., Kaastra J.~S., van der Meer R.~L.~J., Kinkhabwala A., Behar E., Kahn S.~M., 
Paerels F.~B.~S., Sako M., 2002, A\&A, 396, 761 
\bibitem[\protect\citeauthoryear{Canizares et al.}{2005}]{2005PASP..117.1144C} 
Canizares C.~R., et al., 2005, PASP, 117, 1144 
\bibitem[\protect\citeauthoryear{Cash}{1979}]{1979ApJ...228..939C} 
Cash W., 1979, ApJ, 228, 939 
\bibitem[\protect\citeauthoryear{Connolly, McHardy, \& Dwelly}{2014}]{2014MNRAS.440.3503C} 
Connolly S.~D., McHardy I.~M., Dwelly T., 2014, MNRAS, 440, 3503
\bibitem[\protect\citeauthoryear{Crenshaw \& Kraemer}{2007}]{2007ApJ...659..250C} 
Crenshaw D.~M., Kraemer S.~B., 2007, ApJ, 659, 250
\bibitem[\protect\citeauthoryear{Davis et al.}{2014}]{2014ApJ...789..124D} 
Davis B.~L., et al., 2014, ApJ, 789, 124
\bibitem[\protect\citeauthoryear{Guainazzi \& Bianchi}{2007}]{2007MNRAS.374.1290G} 
Guainazzi M., Bianchi S., 2007, MNRAS, 374, 1290 
\bibitem[\protect\citeauthoryear{Guainazzi et al.}{2009}]{2009A&A...505..589G} 
Guainazzi M., Risaliti G., Nucita A., Wang J., Bianchi S., Soria R., Zezas A., 
2009, A\&A, 505, 589 
\bibitem[\protect\citeauthoryear{Hatchett, Buff, \& McCray}{1976}]{1976ApJ...206..847H} 
Hatchett S., Buff J., McCray R., 1976, ApJ, 206, 847
\bibitem[\protect\citeauthoryear{Hunter}{2007}]{2007CSE.....9...90H} 
Hunter J.~D., 2007, CSE, 9, 90 
\bibitem[\protect\citeauthoryear{Kaastra et al.}{2004}]{2004A&A...428...57K} 
Kaastra J.~S., et al., 2004, A\&A, 428, 57 
\bibitem[\protect\citeauthoryear{Kalberla et al.}{2005}]{2005A&A...440..775K} 
Kalberla P.~M.~W., Burton W.~B., Hartmann D., Arnal E.~M., Bajaja E., Morras R., 
P{\"o}ppel W.~G.~L., 2005, A\&A, 440, 775
\bibitem[\protect\citeauthoryear{Kallman \& Bautista}{2001}]{2001ApJS..133..221K}
Kallman T., Bautista M., 2001, ApJS, 133, 221
\bibitem[\protect\citeauthoryear{Kallman et al.}{2014}]{2014ApJ...780..121K} 
Kallman T., Evans D.~A., Marshall H., Canizares C., Longinotti A., Nowak M., Schulz N., 
2014, ApJ, 780, 121 
\bibitem[\protect\citeauthoryear{Kinkhabwala et al.}{2002}]{2002ApJ...575..732K} 
Kinkhabwala A., et al., 2002, ApJ, 575, 732 
\bibitem[\protect\citeauthoryear{Korista, Ferland, \& Baldwin}{1997}]{1997ApJ...487..555K} 
Korista K., Ferland G., Baldwin J., 1997, ApJ, 487, 555
\bibitem[\protect\citeauthoryear{Krolik \& Begelman}{1988}]{1988ApJ...329..702K} 
Krolik J.~H., Begelman M.~C., 1988, ApJ, 329, 702
\bibitem[\protect\citeauthoryear{Landi et al.}{2012}]{2012ApJ...744...99L} 
Landi E., Del Zanna G., Young P.~R., Dere K.~P., Mason H.~E., 2012, ApJ, 744, 99 
\bibitem[\protect\citeauthoryear{Liedahl \& Paerels}{1996}]{1996ApJ...468L..33L} 
Liedahl D.~A., Paerels F., 1996, ApJ, 468, L33 
\bibitem[\protect\citeauthoryear{Madore et al.}{1998}]{1998Natur.395...47M} 
Madore B.~F., et al., 1998, Natur, 395, 47 
\bibitem[\protect\citeauthoryear{Magdziarz \& Zdziarski}{1995}]{1995MNRAS.273..837M} 
Magdziarz P., Zdziarski A.~A., 1995, MNRAS, 273, 837
\bibitem[\protect\citeauthoryear{Maiolino \& Rieke}{1995}]{1995ApJ...454...95M} 
Maiolino R., Rieke G.~H., 1995, ApJ, 454, 95
\bibitem[\protect\citeauthoryear{Maiolino et al.}{2010}]{2010A&A...517A..47M} 
Maiolino R., et al., 2010, A\&A, 517, A47
\bibitem[\protect\citeauthoryear{Matt, Guainazzi, \& Maiolino}{2003}]{2003MNRAS.342..422M} 
Matt G., Guainazzi M., Maiolino R., 2003, MNRAS, 342, 422
\bibitem[\protect\citeauthoryear{Netzer}{2015}]{2015arXiv150500811N} 
Netzer H., 2015, arXiv, arXiv:1505.00811 
\bibitem[\protect\citeauthoryear{Netzer \& Laor}{1993}]{1993ApJ...404L..51N} 
Netzer H., Laor A., 1993, ApJ, 404, L51
\bibitem[\protect\citeauthoryear{Netzer \& Marziani}{2010}]{2010ApJ...724..318N} 
Netzer H., Marziani P., 2010, ApJ, 724, 318
\bibitem[\protect\citeauthoryear{Ogle et al.}{2000}]{2000ApJ...545L..81O} 
Ogle P.~M., Marshall H.~L., Lee J.~C., Canizares C.~R., 2000, ApJ, 545, L81 
\bibitem[\protect\citeauthoryear{Porquet \& Dubau}{2000}]{2000A&AS..143..495P} 
Porquet D., Dubau J., 2000, A\&AS, 143, 495
\bibitem[\protect\citeauthoryear{Risaliti et al.}{2005}]{2005ApJ...630L.129R} 
Risaliti G., Bianchi S., Matt G., Baldi A., Elvis M., Fabbiano G., Zezas A., 2005, ApJ, 630, L129 
\bibitem[\protect\citeauthoryear{Risaliti et al.}{2007}]{2007ApJ...659L.111R} 
Risaliti G., Elvis M., Fabbiano G., Baldi A., Zezas A., Salvati M., 2007, ApJ, 659, L111
\bibitem[\protect\citeauthoryear{Risaliti et al.}{2009}]{2009MNRAS.393L...1R} 
Risaliti G., et al., 2009a, MNRAS, 393, L1 
\bibitem[\protect\citeauthoryear{Risaliti et al.}{2009}]{2009ApJ...696..160R} 
Risaliti G., et al., 2009b, ApJ, 696, 160
\bibitem[\protect\citeauthoryear{Rivers et al.}{2015}]{2015ApJ...804..107R} 
Rivers E., et al., 2015, ApJ, 804, 107
\bibitem[\protect\citeauthoryear{Sako et al.}{2000}]{2000ApJ...543L.115S} 
Sako M., Kahn S.~M., Paerels F., Liedahl D.~A., 2000, ApJ, 543, L115
\bibitem[\protect\citeauthoryear{Schulz et al.}{1999}]{1999A&A...346..764S} 
Schulz H., Komossa S., Schmitz C., M{\"u}cke A., 1999, A\&A, 346, 764
\bibitem[\protect\citeauthoryear{Schurch et al.}{2004}]{2004MNRAS.350....1S} 
Schurch N.~J., Warwick R.~S., Griffiths R.~E., Kahn S.~M., 2004, MNRAS, 350, 1 
\bibitem[\protect\citeauthoryear{Shu, Yaqoob, \& Wang}{2011}]{2011ApJ...738..147S} 
Shu X.~W., Yaqoob T., Wang J.~X., 2011, ApJ, 738, 147
\bibitem[\protect\citeauthoryear{Smith et al.}{2001}]{2001ApJ...556L..91S} 
Smith R.~K., Brickhouse N.~S., Liedahl D.~A., Raymond J.~C., 2001, ApJ, 556, L91 
\bibitem[\protect\citeauthoryear{Stern \& Laor}{2012}]{2012MNRAS.426.2703S} 
Stern J., Laor A., 2012, MNRAS, 426, 2703 
\bibitem[\protect\citeauthoryear{Veilleux, Cecil, \& Bland-Hawthorn}{2005}]{2005ARA&A..43..769V} 
Veilleux S., Cecil G., Bland-Hawthorn J., 2005, ARA\&A, 43, 769
\bibitem[\protect\citeauthoryear{Wang et al.}{2009}]{2009ApJ...694..718W} 
Wang J., Fabbiano G., Elvis M., Risaliti G., Mazzarella J.~M., Howell J.~H., 
Lord S., 2009, ApJ, 694, 718
\bibitem[\protect\citeauthoryear{Wilms, Allen, \& McCray}{2000}]{2000ApJ...542..914W} 
Wilms J., Allen A., McCray R., 2000, ApJ, 542, 914
\end{thebibliography}
\end{document}